\documentstyle[preprint,aps]{revtex}
\begin{document}
\tighten
\title{The $\eta$--meson light nucleus resonances and quasi--bound states}
\author{
S. A. Rakityansky$^{1,2}$, S. A. Sofianos$^1$,  M. Braun$^1$,
        V. B. Belyaev$^2$, and W. Sandhas$^3$}
\address{$^1$ Physics Department, University of South Africa,
 P.O.Box 392, Pretoria 0001, South Africa}
\address{$^2$ Joint Institute  for Nuclear Research, Dubna, 141980,
Russia}
\address{
 $^3$Physikalisches Institut, Universit\H{a}t Bonn, D-53115 Bonn, Germany}
\date{\today}
\maketitle

\begin{abstract}
The position and movement of poles of the amplitude for
elastic $\eta$--meson scattering off the light nuclei $^2H$,\,$^3H$,\,
$^3He$, and $^4He$ are studied. It is found that, within the
existing uncertainties for the elementary $\eta N$ interaction, all these
nuclei can support a quasi--bound state. The values of the $\eta$-nucleus
scattering lengths corresponding to the critical $\eta N$--interaction which
produces a quasi--bound state are given.  \\\\
{PACS numbers: 25.80.-e, 21.45.+v, 25.10.+s}\\\\
\end{abstract}
Since meson factories cannot produce $\eta$--meson beams, these
particles are available for experimental investigations  only as products
of certain nuclear reactions where they appear as final--state particles.
Therefore, final--state interaction effects are the only source of
information about the $\eta$--meson interaction with nucleons.
In this connection,  $\eta$--nucleus systems can play an important role
in investigating the $\eta N$--dynamics, especially if they can form
quasi--bound states. In this case, the final--state
$\eta$--mesons  can be trapped for a relatively long time, and thus the
properties of the $\eta N$--interaction can be studied.\\

Estimations, obtained in the framework of the optical model approach
\cite{haider,liu}, put a lower bound on the atomic number $A$
for which an $\eta$--nucleus bound state could exist, namely $A\ge12$.
In Ref.\cite{li}, the formation of  $\eta$--nucleus states has
been investigated, using the standard Green's function method of
many--body problems. There it was found that an $\eta{^{16}O}$ bound state
should be possible. Experimentally the cross sections of pion-collisions
with lithium, carbon, oxygen, and aluminum, however, gave no evidence
for the existence of $\eta$ bound states with these nuclei \cite{chrien}.\\

A new theoretical analysis of the problem \cite{chiang} predicted
a binding of the $\eta$--meson to $^{12}C$ and heavier nuclei, however, with
rather large widths. The formation of  an  $\eta{^4H}e$ bound state was
studied in a more recent work by Wycech et al.  \cite{wycech},
using a modified multiple scattering theory. These authors obtained a
comparatively large negative value for the real part of the $\eta$--nucleus
scattering length, which was interpreted as an indication that an
$\eta$--nucleus bound state could exist. We note that previous  results
of ours, concerning the $\eta$ scattering lengths with light nuclei
\cite{rak1,rak2,bel1}, showed that the $\eta-{^4H}e$ scattering
length can have an even larger (negative) real part than that of Ref.
\cite{wycech}.\\

In Ref. \cite{rak3}, a preliminary investigation on the possibility
of $\eta$--meson binding in the $d$, $t$, $^3\!He$, and
$^4\!He$ systems was made within the framework of the Finite-Rank
Approximation (FRA) of the nuclear Hamiltonian \cite{bel2,bel3}. The FRA
approach treats the motion of the projectile ($\eta$--meson) and of the
nucleons inside the nucleus separately. As a result the internal dynamics of
the nucleus enters the theory only via the nuclear wave function.
In \cite{rak3}, these wave functions were approximated  by simple Gaussian
forms, which reproduce the nuclear sizes only. In the present work, we
perform calculations with more realistic nuclear wave functions, obtained via
the so--called Integro--Differential Equation Approach (IDEA)
\cite{IDEA1,IDEA2,IDEA3,IDEA4,IDEA5}. We study, in particular,
the position and movement of poles of the elastic amplitude of
$\eta$--meson scattering off the light nuclei $^2H$,\,$^3H$,\,
$^3He$, and $^4He$.\\

The approximate few--body equations in the FRA approach enable us to
calculate the $\eta$--nucleus T--matrix
\begin{equation}
\label{tmat}
     T(\vec{k}',\vec{k};z) \, = \, <\vec{k}',\psi_0|T(z)|\vec{k},\psi_0>\ ,
\end{equation}
at any complex energy. That is, we can locate the poles of the T--matrix
in the complex momentum plane $p=\sqrt{2\mu z}$. Here, $\vec k$ is the
$\eta$--nucleus momentum, $z$  the total energy of the system, $\mu$
the $\eta$--nucleus reduced mass, and $\psi_0$  the nuclear ground-state
wave  function.\\

For the low energies and the light nuclei with only one bound state,
being considered, it appears justified to approximate the target
Hamiltonian $H_A$ by its discrete spectrum
\begin{equation}
\label{appr}
     H_A\approx {\cal E}_0|\psi_0><\psi_0| \ .
\end{equation}
Here $|\psi_0>$ stands for the $^2H$, $^3H$, $^3\!H\!e$, $^4\!H\!e$
bound states, respectively, and ${\cal E}_0$ for the corresponding binding
energies.\\

As a result, we obtain \cite{rak2} for the T--matrix the following
equation
\begin{equation}
\label{t}
       T(z)=\sum\limits_{i=1}^AT_i^0(z)+\sum\limits_{i=1}^AT_i^0(z)
       |\psi_0>\frac{{\cal E}_0}{(z-H_0)(z-H_0-{\cal E}_0)}
       <\psi_0|T(z)\ ,
\end{equation}
where $H_0$ is the $\eta$--nucleus kinetic energy operator, $A$ the number
of nucleons. The $T_i^0(z)$ are Faddeev--type components of
an auxiliary T-operator, which obey the system of coupled  equations
\begin{equation}
\label{t0}
   T_i^0(z)=t_i(z)+t_i(z)\frac{1}{(z-H_0)}\sum\limits_{j\ne i}T_j^0(z)\ .
\end{equation}
Here $t_i$ describes the scattering of the $\eta$--meson off a nucleon at
point $\vec r_i$, where $\vec r_i$ is the  vector from the nuclear center of
mass, which can be expressed in terms of the relative Jacobi vectors
$\{\vec r\}$ of the nucleons. In mixed representation, the operator $t_i$ is
given by
$$
      t_i(\vec k',\vec k;\vec r;z)=t_{\eta N}(\vec k',\vec k;z)
      \exp\left[{{\displaystyle i(\vec k-\vec k')\cdot\vec r_i}}\right],
$$
with $t_{\eta N}(\vec k^{\prime},\vec k;z)$ being the off-shell $\eta N$
amplitude.\\

Thus, to calculate the T--matrix (\ref{tmat}) for any fixed value of the
complex parameter $z=p^2/2\mu$, we have to proceed in three steps. First,
the coupled integral equations
\begin{equation}
\label{t0i}
     T^0_i (\vec{k}',\vec{k};\vec{r};z)
     = t_i(\vec{k}',\vec{k};\vec{r};z) \
  + \int \frac {d^3k''}{(2\pi)^3}
\frac {t_i(\vec{k}',\vec{k}'';\vec{r};z)} {z - \frac {k''^2}{2\mu}}
\sum_{j\neq i} T^0_j(\vec{k}'',\vec{k};\vec{r};z)
\end{equation}
are to be solved for a number of points $\vec r$ in configuration space,
sufficient to perform in a second step the integration
\begin{equation}
              <\vec{k}', \psi_0 |\sum\limits_{i=1}^A T_i^0(z) | \vec{k},
               \psi _0 > = \int d^{3(A-1)}r |
                \psi_0(\vec{r})|^2 \sum\limits_{i=1}^A
                 T_i^0(\vec{k}',\vec{k};\vec{r};z) \label{aver}.
\end{equation}
Having determined these matrix elements, it remains, as a final step, to
solve the integral equation
 \begin{equation}
\label{tm3}
      T(\vec{k}',\vec{k};z) = <\vec{k}',\psi_0|\sum\limits_{i=1}^A T_i^0(z)
      |\vec{k},\psi_0>\, +\  {\cal E}_0\int \frac{d^3k''}{(2\pi)^3}
    \frac{ <\vec{k}',\psi_0|\sum\limits_{i=1}^A T_i^0(z)|\vec{k}'',\psi_0>}{
       (z-\frac{{k''}^2}{2\mu})(z-
        {\cal E}_0-\frac{{k''}^2}{2\mu})}\,T(\vec{k}'',\vec{k};z).
\end{equation}
Note that after partial--wave decomposition both equations (\ref{t0i})
and (\ref{tm3}) become one--dimensional. As an input information, we need
the ground-state wave functions $\psi_0$ of the nuclei involved and the
two-body T--matrix $t_{\eta N}$.\\

The  $\psi_0(\vec r)$ for $A$=3 and $A=4$ were obtained by means of the
IDEA \cite{IDEA1}. In this method
the $A$--body bound state wave function is expanded in Faddeev-type
components,
\begin{equation}
\label{tm4}
     \Psi ({\vec r})=\sum_{i<j\leq A} \,\psi_{ij}({\vec r}),
\end{equation}
given as solutions of
\begin{equation}
    (T-E)\psi_{ij}({\vec r})=-V(r_{ij})\,\sum_{k<l\leq A}
   \,\psi_{kl}({\vec r}),
\end{equation}
where ${\vec r}_{ij}={\vec r}_i-{\vec r}_j$. The IDEA is then introduced
using  the ansatz
\begin{equation}
          \psi_{ij}({\vec r})=
          H_{[L_m]}({\vec r}) P(\zeta_{ij},\rho)/\rho^{(D-1)/2},
\end{equation}
with $\rho=\left[ 2/A\sum r^2_{ij}\right ]^{1/2}$ being the hyperradius,
$D=3(A-1)$, and $H_{[L_m]}({\vec r})$ the harmonic polynomial of minimal
degree $[L_m]$ \cite{Fabr1}. For $[L_m]=0$  the IDEA reads
\begin{equation}
    \left[T+\frac{A(A-1)}{2}V_0(\rho)
   -E\right]\frac{P(\zeta_{ij},\rho)}{\rho^{(D-1)/2}}=
  -[V(r_{ij})-V_0(\rho)]\,\sum_{k<l\leq A}
   \frac{P(\zeta_{kl},\rho)}{\rho^{(D-1)/2}}\ ,
\label{idea1}
\end{equation}
where $V_0(\rho)$ is the so-called hypercentral potential \cite{Fabr1}.
Projecting Eq. (\ref{idea1}) onto the $r_{ij}$--space  provides us,
for spin dependent nucleon-nucleon potentials, with
two coupled integrodifferential equations for the symmetric $S$ and
mixed symmetric $S'$ components of the function $P^n(\zeta_{ij},\rho)$,
$n=S\,\,, S'$. More details and explicit equations are found in
Refs. \cite{IDEA1,IDEA3}.\\

For the nuclear ground states, we use the  fully symmetric S-wave
components obtained with the semi--realistic
Malfliet--Tjon I--III (MT I--III) nucleon-nucleon potential \cite{MTI-III}.
The corresponding two-, three-, and four-body binding energies
are 2.272 MeV, 8.936 MeV, and 30.947 MeV, while the  root
mean square (r.m.s.) radii are 1.976 fm, 1.685 fm and 1.431 fm, respectively.
The omission of Coulomb effects and of the mixed symmetry components makes
$^3H$ and $^3He$ indistinguishable. In order to compensate partly for
this omission, we use in Eq. (\ref{tm3}) the experimental values for
masses and binding energies of the nuclei \cite{Waps}.\\

At low energies the $\eta N$ interaction is dominated by
the $N^*(1535)$   $S_{11}$ - resonance. For the $\eta N$--amplitude we,
therefore, choose the separable form
\begin{equation}
\label{ten}
           t_{\eta N}(k',k;z) = \frac {\lambda}{(k'^2+
           \alpha^2)(z - E_0 + i\Gamma/2)(k^2+\alpha^2)}
\end{equation}
with $E_0 = 1535\; {\rm MeV} - ( m_N + m_{\eta} )$  and $\Gamma = 150\;
{\rm MeV}\;$ \cite{PDGr}. To find the range parameter $\alpha$,
we use the results of Refs. \cite{bhal,Benn}. There the
same $\eta$N $\to$ N$^*$ vertex function ($k^2 + \alpha^2)^{-1}$ was
employed, and $\alpha$ was determined via a two-channel fit to the
$\pi N \to \pi $N and $\pi N \to \eta $N experimental data.\\

Three different values for the  range parameter $\alpha$ are available in
the literature, namely, $\alpha =2.357$~fm$^{-1}$ \cite{bhal}, \,
$\alpha = 3.316$~fm$^{-1}$ \cite{Benn},\, and $\alpha = 7.617$~fm$^{-1} $
\cite{bhal}. Since there is no criterium for singling out  one of them,
we use all three in our calculations. The remaining parameter $\lambda$
is chosen   to provide the correct zero-energy on-shell limit, i.e., to
reproduce the $\eta N$ scattering length $a_{\eta N}$,
\begin{equation}
 t_{\eta  N}(0,0,0) = - \frac {2\pi}{\mu_{\eta  N}}a_{\eta  N}.
\end{equation}
Different analyses provided values for the real part
$Re\,a_{\eta N}$ in the range $0.27 \div 0.98 $ fm and for the imaginary
part $Im\,a_{\eta N}$ in the range $0.19 \div 0.37$ fm \cite{serb}.
To examine at which value of $a_{\eta N}$ within the above ranges an
$\eta$--nucleus bound state exists, we parametrize the scattering length
as follows
\begin{equation}
 \label{length}
       a_{\eta N}=(g0.55+ig'0.30)\,{\rm fm},
\end{equation}
where $g$ and $g'$ are adjustable parameters.\\

Since $a_{\eta N}$ is complex, the $\eta$--nucleus Hamiltonian is
non--Hermitian and its eigenvalues are generally complex. In this case,
eigenvalues attributed to resonances and quasi--bound states are located in
the second--quadrant of the complex $p$--plane \cite{cass}.
The energy $E_0=p^2_0/2\mu$ corresponding to a  pole at $p = p_0$,
\begin{equation}
      E_0=\frac{1}{2\mu}\left[(Re\,p_0)^2-(Im\,p_0)^2+2i
                      (Re\,p_0)(Im\,p_0)\right]\,,
\end{equation}
has a negative real part, $Re\,E_0<0$, only if $p_0$ is above the diagonal
of this quadrant. Such a pole is  related to a quasi--bound state. For
$p_0$ below the diagonal we have $Re\,E_0>0$ and  the pole is  then
attributed to a resonance. Therefore this diagonal is critical:
when crossing it from below a resonance pole becomes a quasi--bound
pole.\\

Fixing  $g$ and $g'$ of Eq. (\ref{length}) to $g=g'=1$ and varying
the complex momentum $p=\sqrt{2\mu z}$, we located the poles close to
the origin $p=0$. The results obtained  are given in Table 1. For
one choice of the range parameter, namely $\alpha=2.357$~fm$^{-1}$,
the positions of the poles  found are shown by the open circles in Fig.1.
It is seen that for the $\eta d$, $\eta t$, and $\eta{^3\!He}$ systems,
these poles lie  in the resonance region, while for the $\eta{^4\!He}$
system the pole is in the quasi--bound region.\\

Increasing  $g$ while keeping $g'=1$, the resonance poles are moving
towards, and finally cross, the diagonal. In the deuteron case, the
corresponding trajectory is depicted in Fig. 1 by the solid curve which
crosses the diagonal when $g=1.6536$. \\

To find the relationship of poles above the diagonal to
physical bound states, we gradually removed the imaginary part of
$a_{\eta N}$  by fixing $g$ and decreasing $g'$ in Eq. (\ref{t0}) to zero.
The imaginary part of the Breit-Wigner factor in Eq. (\ref{ten}) was also
decreased, using the same parameter $g'$, so that it goes over into
$(z - E_0 + i g' \Gamma/2)^{-1}$.
For $g'=0$ the Hamiltonian becomes Hermitian and, hence, the bound state
poles in this case must be on the positive imaginary axis. The dashed curve
in  Fig. 1 is the trajectory of the $\eta d$ bound state pole (with $g=2$)
when $g'$ decreases from 1 to 0. It is seen that the final position of the
pole lies on the  positive imaginary axis. This supports our
interpretation of poles above the diagonal as  quasi-bound states.\\

By varying the enhancing factor $g$ for each of the $\eta$--nucleus systems
under consideration, we found the values which  generate quasi--bound states
on the diagonal. They are given in the Table 2. These values
correspond to an  $\eta N$ attraction, which  generates an $\eta$--nucleus
binding with $Re\,E_0=0$. Further increase of $g$ moves the poles up and
to the right, enhancing the binding and reducing the widths of the states.
The value of $Re \, a_{\eta N}$ which provides the critical  binding lies
within the range $[0.27,0.98]$fm used in the literature.
Therefore, an $\eta$--nucleus quasi--bound states may exist for $A\ge2$.
If this is not the case, then at least near--threshold resonances
(poles just below the diagonal) could exist. However, as can be seen
in Table 1 and 2, the widths of such quasi--bound and resonance states
could be small only for the $\eta{^4\!He}$ system.\\

In  Table 3 we present the $\eta$--nucleus  scattering lengths
calculated with parameters generating the critical $\eta$--nucleus binding.
From this table we see that the real part of the $\eta$--nucleus scattering
length can be small despite the existence of a quasi--bound state.
This is due to the non--Hermitian nature of the $\eta N$ interaction.
Being complex, this interaction generates critical poles rather far from
the origin and their influence on the scattering length (the value of the
amplitude at the origin) is not very strong.\\

In conclusion, we have  shown that the uncertainties in the $\eta N$
scattering length allow for choices of parameters in the
$\eta N$-amplitude that may generate poles in the $\eta$-nucleus
amplitudes considered, which  can be attributed to quasi--bound states.
\acknowledgements{Financial support from the  University of South
Africa  and the Joint Institute  for Nuclear Research, Dubna,
is appreciated.}


\newpage
\noindent
Table 1.\,
Positions of poles  $p_0=\sqrt{2\mu E_0}$ of the $\eta$--nucleus
amplitudes with $g=g'=1$ for the three values of the range parameter $\alpha$.

\begin{center}
\begin{tabular}{|c|c|c|c|}
\hline
 & $p_0\;({\rm fm}^{-1})_{\mathstrut}^{\mathstrut}$ &
             $E_0\;({\rm MeV})$ &
             $\phantom{-}\alpha\;({\rm fm}^{-1})\phantom{-}$\\
\hline\hline
    & $-0.90259+i0.35870$ & $ 31.456-i29.691 $ & 2.357 \\
\cline{2-4}
$\eta\, d$ & $-0.84594+i0.32195$ & $ 28.061 -i24.976 $ & 3.316 \\
\cline{2-4}
    & $-0.82460+i0.30423$ & $ 26.935 -i23.006 $ & 7.617 \\
\hline\hline
    & $-0.56045+i0.23859$ & $ 10.906 -i11.341 $ & 2.357 \\
\cline{2-4}
$\eta\, t$ & $-0.55511+i0.26826$ & $ 10.015 -i12.630 $ & 3.316 \\
\cline{2-4}
    & $-0.51725+i0.27896$ & $ 8.0456 -i12.238 $ & 7.617 \\
\hline\hline
    & $-0.54692+i0.24478$ & $ 10.143 -i11.354 $ & 2.357 \\
\cline{2-4}
$\eta\,{^3\!He}$ & $-0.50815+i0.30402$ & $ 7.0305 -i13.102 $ & 3.316 \\
\cline{2-4}
    & $-0.48310+i0.33948$ & $ 5.0099 -i13.909 $ & 7.617 \\
\hline\hline
    & $-0.16504+i0.27876$ & $-2.0540-i3.7447 $ & 2.357 \\
\cline{2-4}
$\phantom{-}\eta\,{^4\!He}\phantom{-}$ & $-0.20215+i0.38726$ &
                              $-4.4403 -i6.3718 $ & 3.316 \\
\cline{2-4}
    & $\phantom{-}-0.25931+i0.45846\phantom{-}$ &
      $\phantom{-}-5.8175-i9.6766\phantom{-}$ & 7.617 \\
\hline\hline
\end{tabular}
\end{center}
\vspace{5mm}
Table 2.\,
The parameter $g$ generating the $\eta$--nucleus amplitude poles
$p_0=\sqrt{2\mu E_0}$ on the diagonal
for the three values of the range parameter $\alpha$ and $g'=1$.
\begin{center}
\begin{tabular}{|c|c|c|c|c|}
\hline
 & $g$ & $p_0\;({\rm fm}^{-1})_{\mathstrut}^{\mathstrut}$ &
             $E_0\;({\rm MeV})$ &
             $\phantom{-}\alpha\;({\rm fm}^{-1})\phantom{-}$\\
\hline\hline
  & 1.6536  & $-0.32527+i0.32527$ & $-i9.7026$ & 2.357 \\
\cline{2-5}
$\eta\, d$ & 1.5605 & $-0.33541+i0.33541$ & $-i10.317$ & 3.316 \\
\cline{2-5}
    & 1.5260 & $-0.33670+i0.33670$ & $-i10.397$ & 7.617 \\
\hline\hline
    & 1.3624 & $-0.33515+i0.33515$ & $-i9.5266$ & 2.357 \\
\cline{2-5}
$\eta\, t$ & 1.3055 & $-0.35190+i0.35190$ & $-i10.503$ & 3.316 \\
\cline{2-5}
    & 1.2436 & $-0.35186+i0.35186$ & $-i10.500 $ & 7.617 \\
\hline\hline
    & 1.3306 & $-0.34034+i0.34034$ & $-i9.8239$ & 2.357 \\
\cline{2-5}
$\eta\,{^3\!He}$ & 1.2171 & $-0.36267+i0.36267$ & $-i11.155$ & 3.316 \\
\cline{2-5}
    & 1.1421 & $-0.37631+i0.37631$ & $-i12.010$ & 7.617 \\
\hline\hline
    & 0.86222 & $-0.20641+i0.20641$ & $-i3.4679 $ & 2.357 \\
\cline{2-5}
$\phantom{-}\eta\,{^4\!He}\phantom{-}$ & 0.80813 & $-0.26522+i0.26522$ &
                              $-i5.7255 $ & 3.316 \\
\cline{2-5}
    & $\phantom{-}0.79578\phantom{-}$ &
      $\phantom{-}-0.35215+i0.35215\phantom{-}$ &
      $\phantom{-}-i10.094\phantom{-}$ & 7.617 \\
\hline\hline
\end{tabular}
\end{center}
Table 3.\,
The $\eta$--nucleus scattering lengths for the parameter $g$ of Table 2,
which generate the condition for  binding $( {\rm R\!e} E=0 )$.

\begin{center}
\begin{tabular}{|c|c|c|c|}
\hline
             & $\phantom{-}\alpha=2.357\;({\rm fm}^{-1})
             _{\mathstrut}^{\mathstrut}\phantom{-}$ &
             $\phantom{-} \alpha=3.316\;({\rm fm}^{-1})\phantom{-}$ &
             $\phantom{-}\alpha=7.617\;({\rm fm}^{-1})\phantom{-}$\\
\hline
$\eta\, d$ & $0.171+i5.99$ & $ -0.198+i4.57 $ & $-0.318+i3.52 $ \\
\hline
$\eta\, t$ & $-3.65+i3.49$ & $ -2.91+i3.02 $ &$ -2.19+2.70 $\\
\hline
$\eta\,{^3\!He}$ & $-3.49+i3.67$ & $ -2.66+i3.31 $ & $ -1.96+i2.86 $\\
\hline
$\phantom{-}\eta\,{^4\!He}\phantom{-}$ & $-3.43+i2.60$ &
                              $-2.81+i2.14 $ & $-2.30+i1.72 $\\
\hline
\end{tabular}
\end{center}
\def\emline#1#2#3#4#5#6{%
       \put(#1,#2){\special{em:moveto}}
       \put(#4,#5){\special{em:lineto}}}
\def\newpic#1{}
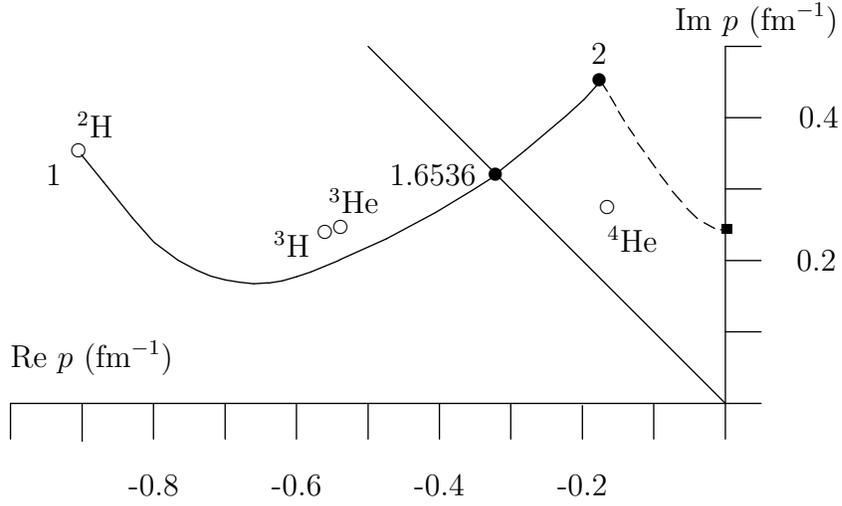
\begin{figure}
\centering
\unitlength=0.95mm
\special{em:linewidth .5pt}
\linethickness{.5pt}
\begin{picture}(170.00,170.00)
\emline{130.00}{80.00}{1}{30.00}{80.00}{2}
\emline{130.00}{130.00}{3}{130.00}{75.00}{4}
\emline{130.00}{80.00}{5}{135.00}{80.00}{6}
\emline{130.00}{90.00}{7}{135.00}{90.00}{8}
\emline{130.00}{100.00}{9}{135.00}{100.00}{10}
\emline{130.00}{110.00}{11}{135.00}{110.00}{12}
\emline{130.00}{120.00}{13}{135.00}{120.00}{14}
\emline{130.00}{130.00}{15}{135.00}{130.00}{16}
\emline{120.00}{80.00}{17}{120.00}{75.00}{18}
\emline{110.00}{79.67}{19}{110.00}{75.00}{20}
\emline{100.00}{80.00}{21}{100.00}{75.00}{22}
\emline{90.00}{80.00}{23}{90.00}{75.00}{24}
\emline{80.00}{80.00}{25}{80.00}{75.00}{26}
\emline{70.00}{80.00}{27}{70.00}{75.00}{28}
\emline{60.00}{80.00}{29}{60.00}{75.00}{30}
\emline{50.00}{80.00}{31}{50.00}{75.00}{32}
\emline{40.00}{80.00}{33}{40.00}{74.67}{34}
\emline{30.00}{80.00}{35}{30.00}{75.00}{36}
\emline{130.00}{80.00}{37}{80.00}{130.00}{38}
\put(112.36,125.25){\circle*{2.00}}
\put(113.50,107.48){\circle{2.00}}
\put(39.47,115.45){\circle{2.00}}
\put(73.98,103.99){\circle{2.00}}
\put(76.14,104.67){\circle{2.00}}
\emline{39.92}{114.78}{39}{42.98}{111.02}{40}
\emline{42.98}{111.02}{41}{47.00}{106.00}{42}
\emline{47.00}{106.00}{43}{50.01}{102.58}{44}
\emline{50.08}{102.53}{45}{53.44}{100.02}{46}
\emline{53.44}{100.02}{47}{56.00}{98.67}{48}
\emline{56.00}{98.67}{49}{58.01}{97.81}{50}
\emline{58.01}{97.81}{51}{60.07}{97.26}{52}
\emline{60.07}{97.26}{53}{61.97}{96.96}{54}
\emline{61.97}{96.96}{55}{64.03}{96.76}{56}
\emline{64.03}{96.76}{57}{66.04}{96.86}{58}
\emline{66.05}{96.84}{59}{68.01}{97.13}{60}
\emline{68.01}{97.13}{61}{69.98}{97.71}{62}
\emline{69.98}{97.71}{63}{72.01}{98.40}{64}
\emline{72.01}{98.40}{65}{76.00}{100.02}{66}
\emline{97.18}{111.53}{67}{101.98}{115.39}{68}
\emline{101.98}{115.39}{69}{107.73}{120.28}{70}
\emline{107.73}{120.28}{71}{110.30}{122.60}{72}
\emline{110.30}{122.60}{73}{111.93}{124.40}{74}
\emline{111.93}{124.40}{75}{112.10}{124.66}{76}
\emline{112.10}{124.66}{77}{112.19}{125.00}{78}
\emline{76.00}{100.07}{79}{82.56}{103.04}{80}
\emline{82.56}{103.04}{81}{89.49}{106.73}{82}
\emline{89.49}{106.73}{83}{94.53}{109.79}{84}
\emline{94.53}{109.79}{85}{97.23}{111.49}{86}
\put(97.86,112.03){\circle*{2.00}}
\emline{112.92}{124.59}{87}{113.50}{123.68}{88}
\emline{113.89}{122.97}{89}{114.67}{121.67}{90}
\emline{115.06}{120.89}{91}{115.91}{119.52}{92}
\emline{116.43}{118.68}{93}{117.27}{117.44}{94}
\emline{117.73}{116.47}{95}{118.64}{115.23}{96}
\emline{119.22}{114.39}{97}{120.26}{112.96}{98}
\emline{120.84}{112.05}{99}{121.95}{110.62}{100}
\emline{122.60}{109.71}{101}{123.90}{108.21}{102}
\emline{124.61}{107.30}{103}{126.11}{105.87}{104}
\emline{127.02}{105.29}{105}{128.51}{104.51}{106}
\emline{129.36}{104.31}{107}{130.01}{104.18}{108}
\put(129.49,103.79){\rule{1.43\unitlength}{1.30\unitlength}}
\put(123.00,133.67){\makebox(0,0)[lc]{Im $p$ (${\rm fm}^{-1}$)}}
\put(140.33,120.00){\makebox(0,0)[lc]{0.4}}
\put(140.00,100.00){\makebox(0,0)[lc]{0.2}}
\put(110.00,70.00){\makebox(0,0)[ct]{-0.2}}
\put(70.00,70.00){\makebox(0,0)[ct]{-0.6}}
\put(90.00,70.00){\makebox(0,0)[ct]{-0.4}}
\put(50.00,70.00){\makebox(0,0)[ct]{-0.8}}
\put(30.00,85.00){\makebox(0,0)[lb]{Re $p$ (${\rm fm}^{-1}$)}}
\put(36.00,112.00){\makebox(0,0)[cc]{1}}
\put(44.33,119.00){\makebox(0,0)[rc]{$^2{\rm H}$}}
\put(69.33,102.33){\makebox(0,0)[cc]{$^3{\rm H}$}}
\put(78.00,108.40){\makebox(0,0)[cc]{$^3{\rm He}$}}
\put(83.00,112.00){\makebox(0,0)[lc]{1.6536}}
\put(112.33,129.00){\makebox(0,0)[cc]{2}}
\put(117.00,103.00){\makebox(0,0)[cc]{$^4{\rm He}$}}
\end{picture}
\caption{ The $\eta$--nucleus elastic scattering amplitude pole
positions in the complex $p$--plane. The open circles correspond to $g$=1.
The solid curve is the $\eta d$--amplitude pole trajectory
when $g$ increases from $g$=1 to $g$=2. The dashed curve shows
the trajectory of the $\eta d$
pole with $g$=2 and with $g'$ varied until
the $\eta N$ interaction becomes real.}
\label{ttt:pic}
\end{figure}

\end{document}